\documentclass{vldb}

\pdfoutput=1
\usepackage[usenames]{color}
\usepackage{setspace}
\usepackage{xspace}
\usepackage{graphicx}
\usepackage{bm}
\usepackage{amsmath}
\usepackage{amssymb}
\usepackage{listings}
\usepackage{algorithm}
\usepackage{algpseudocode}
\usepackage{multirow}
\usepackage{braket}
\usepackage{enumitem}
\usepackage{caption}
\usepackage[font=bf, labelfont=bf]{caption}
\usepackage{subcaption}
\usepackage[font=small,labelfont=small]{subcaption}

\usepackage[T1]{fontenc}
\usepackage[utf8]{inputenc}
\usepackage{authblk}

\newlist{inlineroman}{enumerate*}{1}
\setlist[inlineroman]{itemjoin*={{, and }},afterlabel=~,label=\roman*.}

\newcommand{\Commentblock}[2]{{\color{blue}{#1} :::  \color{red}{#2}}\\}

\newcommand{\pluseq}{\mathrel{+}=}

\begin{document}

\title{ {\large \textbf{\textit{q}H\textit{i}PSTER: The Quantum High Performance Software Testing Environment}} }

\numberofauthors{3} 

\author[1]{Mikhail Smelyanskiy}
\author[1,2]{Nicolas P. D. Sawaya}
\author[2]{Al\'{a}n Aspuru-Guzik}
\affil[1]{Parallel Computing Lab, Intel Corporation}
\affil[2]{Department of Chemistry and Chemical Biology, Harvard University}
\date{\today}

\maketitle

\begin{abstract}
We present \textit{q}H\textit{i}PSTER, the Quantum High Performance Software Testing Environment. \textit{q}H\textit{i}PSTER is a distributed high-performance implementation of a quantum simulator on a classical computer, that can simulate general single-qubit gates and two-qubit controlled gates. We perform a number of single- and multi-node optimizations, including vectorization, multi-threading, cache blocking, as well as overlapping computation with communication. Using the TACC Stampede supercomputer, we simulate quantum circuits (``quantum software") of up to 40 qubits. We carry out a detailed  performance analysis to show that our simulator achieves both high performance and high hardware efficiency, limited only by the sustainable memory and network bandwidth of the machine.
\end{abstract}

\section{Introduction}

Simulation has long been an invaluable tool for modeling classical computer systems, such as digital circuits, processor microarchitecture, or interconnection networks. For example, processor designers rely on cycle-accurate simulators to characterize the impact of new hardware features on application performance and CPU power~\cite{Carlson:2011:SEL:2063384.2063454,Li:2009:MIP:1669112.1669172}. System designers use network simulators to quantify the impact of new network topologies on communication pattern of a distributed applications~\cite{Rodrigues:2011:SST:1964218.1964225}.

Similarly, using a classical computer to simulate a quantum system is important for better understanding its behavior. Such simulations can be used to validate the complexity of new quantum algorithms, to study quantum circuits that are difficult to characterize analytically, or to investigate the performance of circuits in the presence of noise.

For example, many algorithmic choices must be made in the design of a quantum circuit for calculating molecular energies \cite{aag2005,peruzzomcclean2014,mcclean2015}. There are multiple methods for mapping the problem onto a set of qubits\cite{whitfield2011,love2012}, as well as many possible gate sequences for approximating the operators \cite{Hastings2015,babbush2015}. Using a high performance simulator, one can study the effects of these parameters on the algorithm's performance. This in turn helps to minimize the quantity of quantum resources required to bound the error below a certain value. Alternatively, one may wish to implement various noise models to determine the minimum required decoherence times for a given level of accuracy. Simulating other proposed quantum algorithms, including the Hubbard model\cite{wecker2015} and the finite element method\cite{clader2013}, might provide similar insight.

While there exists number of techniques to simulate specific classes of quantum circuits efficiently~\cite{2003RSPSA.459.2011J,Markov:2008:SQC:1405087.1405105,2006quant.ph.11156A,2003PhRvL..91n7902V,1998quant.ph..7006G},  simulation of generic quantum circuits  on classical computers is very inefficient, due to the exponential overhead~\cite{Feynman82simulatingphysics}. Specifically, the fundamental challenge is that the size of state, or number of quantum amplitudes,  grows exponentially with the number of qubits. Given  $n$ qubits, the size of the state vector is $2^n$ complex amplitudes, or $2^{n+4}$ bytes.\footnote{Here and in the rest of the paper we assume complex double precision, with eight-byte real and eight-byte imaginary parts.} Thus, the memory capacity of the classical system  imposes an upper bound on the size of the simulation.  In addition, the size of the quantum circuit (its number of gates) can result in significant run-time requirements on the classical system.

\begin{table}[]
\centering

\begin{tabular}{|r|c|c|}
\hline
    \textbf{System}  & \textbf{Memory (PB)}     & \textbf{Max qubits}          \\ \hline
TACC Stampede        & 0.192   & 43  \\
Titan                  & 0.71   &    45 \\
K computer      & 1.4    &  46     \\
APEX2020~\cite{APEX}        & 4-10    &  48-49\\
\hline
\end{tabular}
\vspace{2mm}
\caption{Examples of TOP500 supercomputing systems, their memory  capacity (in Petabytes), and largest  quantum system they can simulate.}
\label{table:hpcvsqubits}
\end{table}

These challenges can be mitigated by taking advantage of high-performance distributed computation. Most existing quantum simulators run on a single CPU~\cite{QuantumSimulators}. Such simulations are limited to $ \approx 30 - 33$ qubits  due to limited memory capacity and bandwidth of a single node. Several distributed quantum simulators were developed which could simulate more qubits, compared to a single node, due to larger aggregate memory capacity of the system. One of the first distributed quantum simulators, developed in 2002, simulated up to 30 qubits on the Sun Enterprise 4500 system~\cite{GPQCsim}. More recently, another simulator, called JUMPIQCS,   developed to model impact of noise on quantum simulation, simulated 36 qubits on the JUMP IBM p690+ system~\cite{DBLP:phd/de/Trieu2010}. 

Table~\ref{table:hpcvsqubits} shows examples of several supercomputers among the top ten systems in the most recent TOP500 
list\footnote{TOP500~\cite{top500} ranks 500 most powerful commercially available supercomputers in the world, based on the scoring from  LINPACK~\cite{HPL} benchmark. The list is  compiled twice a year; the most recent compilation occurred in November, 2015.}, their aggregate memory capacity, and the number of qubits the system can simulate. We see that today's practical limit is 46 qubits. $48-49$ qubit simulations will become possible in 2020 with the arrival of NERSC-9 and Crossroads pre-exascale systems~\cite{APEX}.

While the aggregate capacity of a particular HPC system is fixed, the quantum simulation time can be further improved. Herein we describe the implementation of \textit{q}H\textit{i}PSTER and the optimization required to achieve high performance and high hardware efficiency on the Stampede supercomputer. Using 1024 nodes,the  maximum available allocation, we simulate quantum circuits of up to 40 qubits. For a 40-qubit system, when no communication is required, single- and two-qubit controlled gate operations are memory bandwidth bound and take $0.43$ and $0.21$ seconds, respectively. Cache blocking optimization results in an additional $\approx2.56\times$ run-time reduction of these gate operations. When communication is required, these gate operations becomes network bandwidth bound, and their run-time increases by $10\times$, which is commensurate with the memory to network bandwidth ratio on Stampede. Finally, using 1024-node distributed simulation, we simulate the 40-qubit quantum Fourier transform, an important kernel of many quantum algorithms, in 997 seconds.

Section~\ref{sec:background} briefly reviews single- and two- qubit gate operations. Section~\ref{sec:Implementation} describes the single and multi-node implementations of both of these operations in \textit{q}H\textit{i}PSTER. Section~\ref{sec:archalgopt} discusses  architectural and algorithmic optimizations, while Section~\ref{sec:performance} presents experimental results and detailed performance analysis of \textit{q}H\textit{i}PSTER. Future  directions to further improve simulator performance are described in Section~\ref{sec:futurework}.

\section{Background}
\label{sec:background}

The current version of \textit{q}H\textit{i}PSTER propagates only pure states. Hence, we operate on the $2^N\times1$ state vector instead of the $2^N\times2^N$ density matrix~\cite{MikeIke}. This approach does not preclude the simulation of mixed states when studying the effects of noise and gate errors. This is due to the fact that there are methods for reconstructing a density matrix from many iterations of simulated pure states \cite{bassi08}.

We focus on implementing general single-qubit gates as well as two-qubit controlled gates (including, \textit{controlled-NOT} gate), which are known to be universal~\cite{PhysRevA.51.1015}. A quantum single-qubit gate operation on qubit $k$ can be represented by a unitary transformation:
\begin{equation}
\label{eq:fullunitarymatrix}
U =  I \otimes I \otimes ... \otimes Q \otimes ... \otimes I  \otimes I
\end{equation}
 where $Q$ is 2x2 unitary matrix,
\[ 
Q=\left(
\begin{array}{cc}
q_{11} & q_{12} \\
q_{21} & q_{22} \\
\end{array} 
\right)
\]

\begin{figure}
    \centering
    \begin{subfigure}[b]{0.14\textwidth}
        \includegraphics[width=\textwidth]{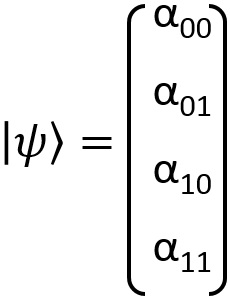}
        \caption{Quantum state}
        \label{fig:qstate}
    \end{subfigure}
    \hspace{4mm}
    \begin{subfigure}[b]{0.28\textwidth}
        \includegraphics[width=\textwidth]{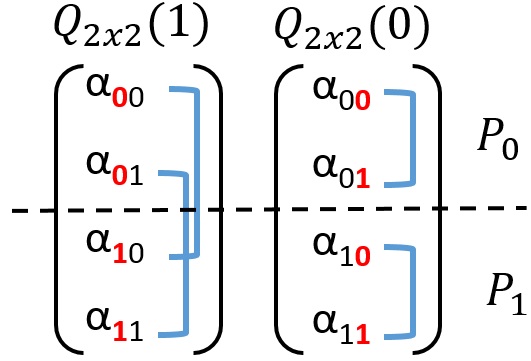}
        \caption{Single-qubit gate}
        \label{fig:gate}
    \end{subfigure}
    \caption{Example of (a) two-qubit quantum state, and (b) single-qubit gate operations, applied to qubits $0$ and $1$, respectively. Subscripts shows the binary representation of the state amplitude index. (b) also shows an example of  distributed computation of gate operation on two processors, $P_0$ and $P_1$.} \label{fig:examplequantgate}
\end{figure}

However, one does not need to construct the entire $U$, and can perform $Q$ transformation directly on the state vector, as shown in Figure~\ref{fig:examplequantgate}, using an example of two qubits. Figure~\ref{fig:examplequantgate}(a) shows the vector representation of a quantum state. Each amplitude has a subscript index in the binary representation. Figure~\ref{fig:examplequantgate}(b) shows single-qubit gate operations on qubit 0 and 1, respectively. Applying a single-qubit gate to qubit $0$  is equivalent to applying $Q$ to every pair of amplitudes, whose indices have $0$ and $1$ in the first bit, while all other bits remain the same. Similarly, applying single-qubit gate to qubit $1$ applies $Q$ to every pair of amplitudes whose indices  differ in their second bit. More generally, performing a single-qubit gate on qubit $k$ of $n$-qubit quantum register applies $Q$ to pairs of amplitudes whose indices differ in $k$-th bits of their binary index:

\begin{equation}
\begin{aligned}
\alpha'_{*...*0_{k}*...*} = q_{11} \cdot \alpha_{*...*0_{k}*...*} + q_{12} \cdot \alpha_{*...*1_{k}*...*} \\
\alpha'_{*...*1_{k}*...*} = q_{21} \cdot \alpha_{*...*0_{k}*...*} + q_{22} \cdot \alpha_{*...*1_{k}*...*}  
\label{eq:sqgunitaryoperation}
\end{aligned}	
\end{equation}

When the state vector is dense and  stored sequentially in memory, the stride between $\alpha_{*...*0_{K}*...*}$ and $\alpha_{*...*1_{k}*...*}$ is $2^k$. For example, in Figure~\ref{fig:examplequantgate}(b), the quantum gate applied to qubit $0$ results in a stride of $1$ ($2^0$), while operating on qubit $1$ results in stride of $2$ ($2^1$).  Gate operations applied to high-order qubits result in large strides, and, as the result, pose  challenges both to single and distributed implementations, as described in the later sections.

A generalized two-qubit \textit{controlled-Q} gate, with  a control qubit $c$ and a target qubit $t$, works as follows: if $c$ is set to $\ket{1}$, $Q$ is applied to $t$; otherwise $t$ is left unmodified:

\begin{equation}
\begin{aligned}
\alpha'_{*1_{c}*0_{t}*...*} = q_{11} \cdot \alpha_{*1_{c}*0_{t}*...*} + q_{12} \cdot \alpha_{*1_{c}*1_{t}*...*} \\
\alpha'_{*1_{c}*1_{t}*...*} = q_{21} \cdot \alpha_{*1_{c}*0_{t}*...*} + q_{22} \cdot \alpha_{*1_{c}*1_{t}*...*}  
\label{eq:cqgunitaryoperation}
\end{aligned}	
\end{equation}

\section{Implementation}
\label{sec:Implementation}

This section describes the \textit{q}H\textit{i}PSTER implementation of single and controlled-Q gates for both single as well as multiple nodes.

\subsection{Single node implementation}
\label{subsec:singlenoderimpl}

The single node implementation of a single-qubit gate is trivial, and directly follows from Equation~\ref{eq:sqgunitaryoperation}, as shown in Figure~\ref{fig:sqgpseudocode} for an $n$-qubit quantum system. Namely, the outer loop iterates over consecutive groups of amplitudes of length $2^{k+1}$, while inner loop applies $Q$ to every pair of amplitudes within the group, separated by the stride of $2^k$.  The controlled-$Q$ operation is similar, except that it requires another (outer) loop  to skip over the amplitudes which correspond to the controlled qubit of $\ket{0}$ (or equivalently, a $c$-th bit of 0 in the binary representation of amplitude's index). 

\begin{figure}[t!]
\begin{small}
\begin{algorithmic}[1]
\For{$g \leftarrow 0; g < 2^n; g \pluseq 2^{k+1} $}  
  \For{$i \leftarrow g; i < g+2^k; i++ $} 
  \vspace{2mm}
    \State $\alpha'_{i} \leftarrow q_{11} \cdot \alpha_{i} + q_{12} \cdot \alpha_{i+2^K} $\\
    \vspace{-1.5mm}
    \State $\alpha'_{i+2^K} \leftarrow q_{21} \cdot \alpha_{i} + q_{22} \cdot \alpha_{i+2^K} $\\
  \EndFor
\EndFor
\end{algorithmic}
\end{small}
\vspace{2mm}
\caption{
Sinlge-qubit gate operation pseudo-code.
}
\label{fig:sqgpseudocode}
\end{figure}

\begin{figure}
\centering
\includegraphics[width=0.5\columnwidth]
{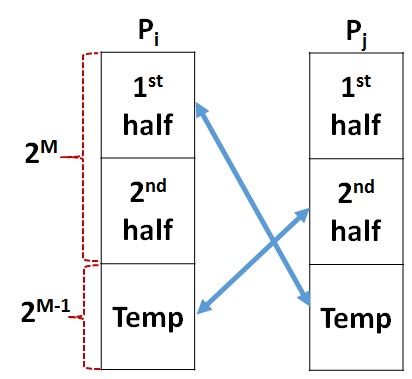}
\vspace{2mm}
\caption{Distributed implementation of a single-qubit gate operation on qubit $k$. Communication occurs between pairs of processors, $2^{k-m}$ apart.  Each processor uses temporary storage to hold half of the state from the other processor. Processors exchange half of their states, compute on exchanged halves, and then perform another exchange.}
\label{fig:distrimpl}
\end{figure}

\subsection{Distributed Implementation}
\label{subsec:distrimpl}

In our distributed implementation, a state vector of $2^n$ amplitudes ($2^{n+4}$ bytes) is distributed among 
$2^p$ nodes, such that each node stores a local state of $2^{n-p}$ amplitudes. Let $m=n-p$. Naturally, $2^{m+4}$ must be less than the total memory capacity of the node. 

Given single-qubit gate operation $Q$ on qubit $k$, if $k < m$, the operation is fully contained within a node. When $k \geqslant m$, the first and second elements of the pair are located on two different nodes and communication is required. Note that the distance between communicating processors in the virtual topology is $2^{k-m}$. We implement the communication scheme described  in~\cite{DBLP:phd/de/Trieu2010}. Our enhancements to this scheme are described in Section~\ref{sec:archalgopt}. Figure~\ref{fig:distrimpl} demonstrates the scheme. Given a local state vector of $2^m$ complex amplitudes, each node reserves an extra $2^{m-1}$ words of memory as temporary storage. Each local state vector is logically partitioned into two halves. Both nodes perform pairwise exchange of these halves: $P_i$ sends its first half to $P_j$, while $P_j$ sends its second half to $P_i$. Each node places the received half  into its own temporary storage.  Next, $P_i$ applies $Q$ to its first half and the temporary storage (which contains first half of $P_j$), while $P_j$ applies $Q$ to its second half and the temporary storage (which contains second half of $P_j$). This results in $P_i$ updating $P_j$'s second half, and $P_j$ updating $P_i$'s first half. This is followed by another pair-wise exchange, where $P_i$ sends $P_j$'s updated half back to $P_j$, while  $P_j$ sends $P_i$'s updated half to $P_j$. This completes the distributed state update. The advantage of this approach is that it distributes work evenly among pairs of nodes.

The distributed implementation of controlled gate operation with controlled qubit $c$ and target qubit $t$ is more involved. 
If $t < m$, there is no communication, while for $t \ge m$ communication is required. In addition, for each of these two cases, we use different kernels, depending on whether $c < m$, or $c \ge m$. Therefore, there are total of four cases for a controlled qubit gate, as opposed to only two for single-qubit gate. We omit further details for brevity. 

\section{Architectural and Algorithmic Optimization}
\label{sec:archalgopt}

In this section we describe several single- and multi-node performance optimizations  we apply to achieve high performance in \textit{q}H\textit{i}PSTER.

\subsection{Vectorization}
\label{subsec:vectorization}

Both single and controlled gate operations have the same inner loop shown in Figure~\ref{fig:sqgpseudocode} (Lines 2-7). This loop is data parallel: every iteration performs the same operation on different set of data. One of the most common and energy efficient methods for exploiting data-level parallelism is via single-instruction-multiple-data (SIMD) execution.  In SIMD execution, a single instruction operates on multiple data elements simultaneously.  This is typically implemented by extending the width of registers and ALUs, allowing them to hold or operate on multiple data elements, respectively.  The rest of the system is left untouched. 

Modern Intel CPUs support SIMD (Single Instruction Multiple Data) instructions, such as AVX2~\cite{AVX2}, which can perform 4 double-precision operations simultaneously on 4 elements of the input registers. We map every two  iterations of the inner loop of Figure~\ref{fig:sqgpseudocode} to 4-wide SIMD instructions; each iteration, which operates on a complex number composed of real and imaginary parts, is mapped to two entries of the SIMD register. We developed specialized code, using compiler intrinsics, to efficiently perform complex arithmetic using SIMD instructions. 

\subsection{Threading}
\label{subsec:threading}

Modern CPUs have multiple cores, and some have several threads per core. To achieve good performance it is important to  parallelize the workload among these threads. There are two levels of parallelization of the code in Fig.~\ref{fig:sqgpseudocode}: one over the inner loop and the other over the outer loop\footnote{Controlled gates introduce an additional loop nest and thus an additional level of parallelization}. Note that the outer loop performs $2^{n-k-1}$ iterations, while the inner loop performs $2^k$ iterations; a smaller $k$ results in more (less) outer (inner) loop iterations, while larger $k$ has the opposite effect. For example, when $k=n-4$, the outer loop only performs eight iterations, which will leave some cores idle when this loop is parallelized on the CPU with more than eight threads. To choose which loop to parallelize, we dynamically check the number of iterations and parallelize at the nest level with the largest amount of work.
 
\subsection{Improvement in Communication}
\label{subsec:communication}

We improve upon the communication scheme described in Section~\ref{sec:Implementation}, which requires an extra $2^{m-1}$ words of temporary storage per node. This implementation is wasteful. Consider, for example, a node with 48GB of main memory. In theory, one could use 32GB for state vector and remaining 16GB for temporary storage to simulate system with 31 qubits. In practice, no  application can use the entire memory of the machine without significant performance impact due to paging, because some of the memory is reserved for the OS, system software, and other purposes. Hence we can only simulate quantum system with 30 qubits in this case. 

To reduce the memory requirements of temporary storage, we divide the distributed phase into multiple steps. At each step we exchange and reserve temporary storage for only a small portion of the state vector, as opposed to the entire half as in the original approach. We apply gate operations only to this portion. Each steps reuses the same temporary storage, which is dramatically reduced. Continuing with the previous example, we can use 8GB of temporary storage instead of 16GB. It requires two steps, but enables simulating 31 qubits - one qubit more than with our original approach. As long as the amount of data exchanged within each step is large enough to saturate the network bandwidth, the overall run-time remains the same as in original approach.

The total run-time of the distributed application, $T_{tot}$, is a function of compute time, $T_{comp}$, and communication time, $T_{comm}$. The original implementation of the quantum simulator performs these phases separately, resulting in $T_{tot} = T_{comp} + T_{comm}$. Using a multistep approach, described above, we overlap communication and computation in step $i$ with state exchange in steps $i-1$ and $i+2$. This results in $T_{tot} = max(T_{comp}, T_{comm})$, thus partially hiding overhead of communication.

\subsection{Cache Blocking  through Gate Fusion}
\label{subsec:fusion}

\begin{figure}[b!]
\begin{small}
\begin{algorithmic}[1]
\For{$gb \leftarrow 0; gb < 2^n; gb \pluseq 2^c $}  
  \For{group of gates, on some qubit $k$ ($k < {l_c}$)} 
    \For{$g \leftarrow gb; g < gb + 2^{l_c}; g \pluseq 2^{k+1} $}  
    \For{$i \leftarrow g; i < g+2^k; i++ $} 
    \vspace{1mm}
      \State $\alpha'_{i} \leftarrow q_{11} \cdot \alpha_{i} + q_{12} \cdot \alpha_{i+2^K} $\\
      \State $\alpha'_{i+2^K} \leftarrow q_{21} \cdot \alpha_{i} + q_{22} \cdot \alpha_{i+2^K} $\\
    \EndFor
    \EndFor
  \EndFor
\EndFor
\end{algorithmic}
\end{small}
\vspace{2mm}
\caption{
Pseudo-code shown gate fusion performed on a block of fused gates.}
\label{fig:sqgfusionpseudocode}
\end{figure}

Single and controlled qubit operations perform small amounts of computation. Therefore,  their performance is limited by memory bandwidth when the size of the state vector exceeds the size of the Last Level Cache (LLC). Modern CPUs have large LLCs, tens of MBytes per socket. LLC has a much higher bandwidth than memory (albeit, much smaller capacity), but taking advantage of high LLC bandwidth requires restructuring  the algorithm so that its working set fits in an LLC~\cite{Lam:1991:CPO:106973.106981}.

Using fused gates we can block computation in LLC as shown in Figure~\ref{fig:sqgfusionpseudocode}. Assume that LLC has a size of $2^{l_c}$. For a given quantum circuit, we identify groups of consecutive gates, where each gate operates on some qubit $k$,   $k < {l_c}$.  We iterate over blocks of $2^{l_c}$ amplitudes of the state vector (Line 1). Each of the fused gates is applied to this block (Lines 2-4), while the block remains resident in  LLC and therefore can benefit from the LLC's high bandwidth. 

\section{Performance}
\label{sec:performance}

\subsection{Experimental Setup}
\label{subsec:setup}

\begin{sloppypar}
We evaluate the performance and scalability of \textit{q}H\textit{i}PSTER on the Stampede supercomputer. Stampede~\cite{Stampede} at the Texas Advanced Computing Center (TACC)/Univ. of Texas, USA (\# 10 in the current TOP500 list) consists of 6,400 compute nodes, each of which is equipped with two sockets of Xeon E5-2680  connected via QPI and 32GB of DDR4 memory per node (16GB per socket), as well as one Intel\textregistered  Xeon Phi\texttrademark  SE10P co-processor. Each socket has 8 cores, with hyperthreading disabled. We use OpenMP~4.0~\cite{openmp13} to parallelize computation among threads. The nodes are connected via a Mellanox FDR 56 Gb/s InfiniBand interconnect. Both sockets within the node share a single network card, connected via PCIe. Therefore, when both sockets communicate at the same time, each socket  only gets half of the available network injection bandwidth. 
\end{sloppypar}

In this evaluation we have used only 1000 nodes (2000 sockets), the maximum available allocation, and did not use Xeon Phi  co-processors. With aggregate memory capacity of 32 Tbytes across 1000 nodes, we were able to simulate quantum system of up to 40 qubits.

\begin{table}[]
\centering

\begin{tabular}{|c|l|c|c|}
\hline
\textbf{Case} & \textbf{Operation}              & \textbf{Analytic}  & \textbf{Stampede}  \\\hline 
& Single-qubit  gate                &   &    \\
1 &\hspace{0.2cm} $k < m$        &  $\frac{2^{m+5}}{B_{mem}}$  & 0.43 sec \\[0.13cm]
2 & \hspace{0.2cm} $k \ge m$      & $\frac{2^{m+5}}{B_{net}}$  &  3.12 sec \\[0.13cm] \hline
& Two-qubit gate                 &      &         \\
3 & \hspace{0.2cm} $t < m, c < m$   & $\frac{2^{m+4}}{B_{mem}}$   &  0.21 sec           \\[0.13cm]
4 & \hspace{0.2cm} $t < m, c \ge m$   & $\frac{2^{m+5}}{B_{mem}}$     &  0.43  sec          \\[0.2cm]
5 & \hspace{0.2cm} $t \ge m, c < m$   & $\frac{2^{m+4}}{B_{net}}$      &  1.56   sec        \\[0.13cm]
6 & \hspace{0.2cm} $t \ge m, c \ge m$   & $\frac{2^{m+5}}{B_{net}}$     &  3.12    sec        \\[0.09cm]
\hline
\end{tabular}
\vspace{2mm}
\caption{Lower bound (`best case') time per gate for single and two-qubit gates. The first column shows different cases. The second column shows the analytic expression, as the function of $m$, $B_{mem}$, and $B_{net}$. The third column shows specific times on our experimental platform for $n=29$, $B_{mem}=40$ GB/s, and $B_{net}=5.5$ GB/s.}
\label{table:perfexpectations}
\end{table}

\subsection{Lower Bounds on Achievable Performance} 
\label{subsec:lowerbounds}
 
The run-time of data intensive application, such as a quantum circuit simulator, is bound either by memory bandwidth, when run on single node, or by network bandwidth, when run in distributed fashion. The memory bound is equal to the total required memory traffic divided by the sustainable memory bandwidth ($B_{mem}$),  measured by STREAM Copy benchmark~\cite{McCalpin2007}. The network bound is equal to the total amount of network traffic divided by sustainable network bandwidth ($B_{net}$), measured by the OSU bandwidth benchmark~\cite{OSU}. On our system, $B_{mem}=40$ GB/s, while the $B_{net}=5.5$ GB/s (bidirectional), per socket. The closer the actual application run-time is to one of those two bounds, the higher its hardware efficiency is. 

\

Table~\ref{table:perfexpectations} shows lower run-time bounds for single and controlled qubit gates, with and without communication. For example, the expected lower bound  for single-qubit operation  when no communication is required ($k < m$, Case 1) is $2^{m+5} / B_{mem}$ seconds. Here,  $2^{m+4}$ is the size of the state vector in bytes, while an additional factor of two is due to the fact that the state vector is both read and written, which doubles the amount of memory traffic. Since our system has 16GB of memory per socket, it can simulate at most 29 qubits within a socket, because the state vector occupies $8.5$GB of memory. With stream bandwidth $B_{mem}=40$ GB/s,  single socket memory bound is 0.43 seconds for 29-qubit quantum simulation. Since the controlled qubit gate accesses only half of the state vector, its expected runtime is 0.22 seconds, also when there is no communication (Case 3). When communication is required, a pair of nodes performs two exchanges of half of the state with another node, as described in Section~\ref{subsec:distrimpl}. This results in network bound of $(2^{m+5}) / B_{mem}$. For $m=29$, the corresponding network bound is 3.12 seconds, which is $7.3\times$ higher than memory bound, and is effectively a ratio between memory and network bandwidth.

In the remainder of the section we use these bounds to explain the results of our experiments.

\subsection{Single Node Performance}
\label{subsec:singlenodeperf}
 
\begin{figure}
\centering
\includegraphics[width=1.1\columnwidth]
{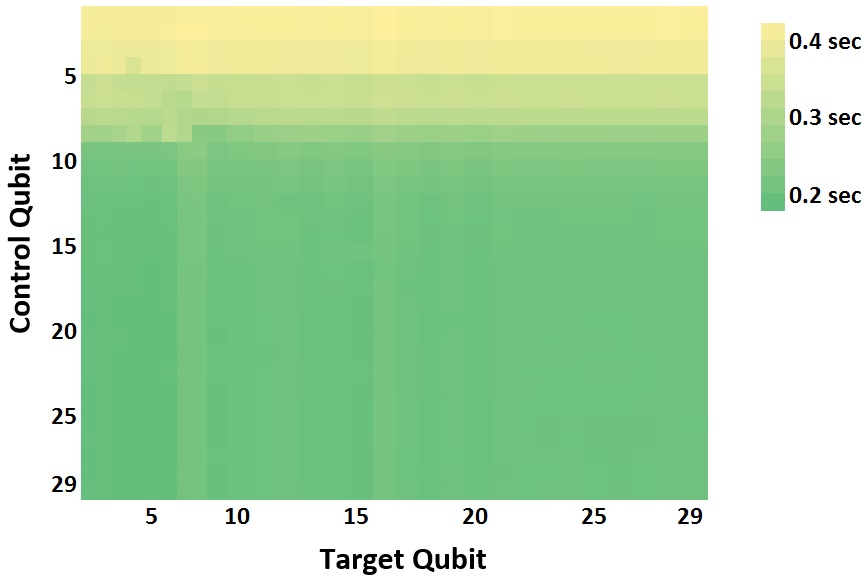}
    \vspace{1mm}
\caption{Performance of controlled single-qubit gate on a single node for a 29-qubit system ($n = 29$)}

\label{fig:cgheatmap}
\end{figure}

Our single-qubit gate operation runs very close to memory bound, regardless of which qubit the gate is applied to. Figure~\ref{fig:cgheatmap} shows a runtime heatmap of two-qubit controlled gate operations for 29 qubits ($n=29$), and all combinations of control and target qubits. Green corresponds to the high end of the performance spectrum, that is memory bound of 0.21 seconds (Case 3,  Table~\ref{table:perfexpectations}). We see that low values of the control qubits result in suboptimal performance of 0.4 seconds ($\approx 2\times$ higher than memory bound). As the value of control qubit increases, the performance improves, and starting from control qubit 10 it approaches memory bound. The reason for suboptimal performance at a lower number of qubits is as follows. Recall that a control gate affects only the amplitudes whose $c$'th bit is set to  1. Thus, the memory access pattern only accesses the second half of every $2^c$-long memory region. These 'holes' break the stride and interrupt the hardware prefetcher which is trained to prefetch elements which are at a constant stride from each other~\cite{inteloptimize}. Not only does the failure of the prefetcher to detect the stride  expose cache miss latency, it also results in wasted memory bandwidth, due to prefetching useless data. This  results in run-time increase.

\begin{figure}
\centering
\includegraphics[width=0.9\columnwidth]
{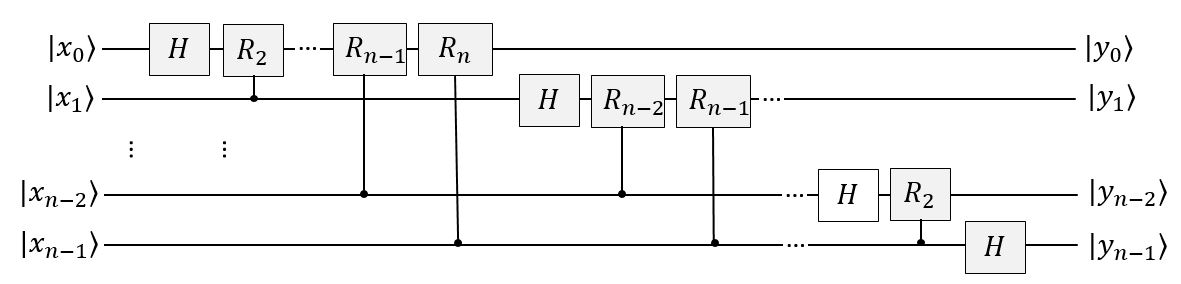}
    \vspace{0.1mm}
\caption{{Inverse Quantum Fourier Transform} circuit applied to an $n$-qubit quantum register.}
\label{fig:iqfft}
\end{figure}

\begin{figure}
\centering
\includegraphics[width=0.98\columnwidth]
{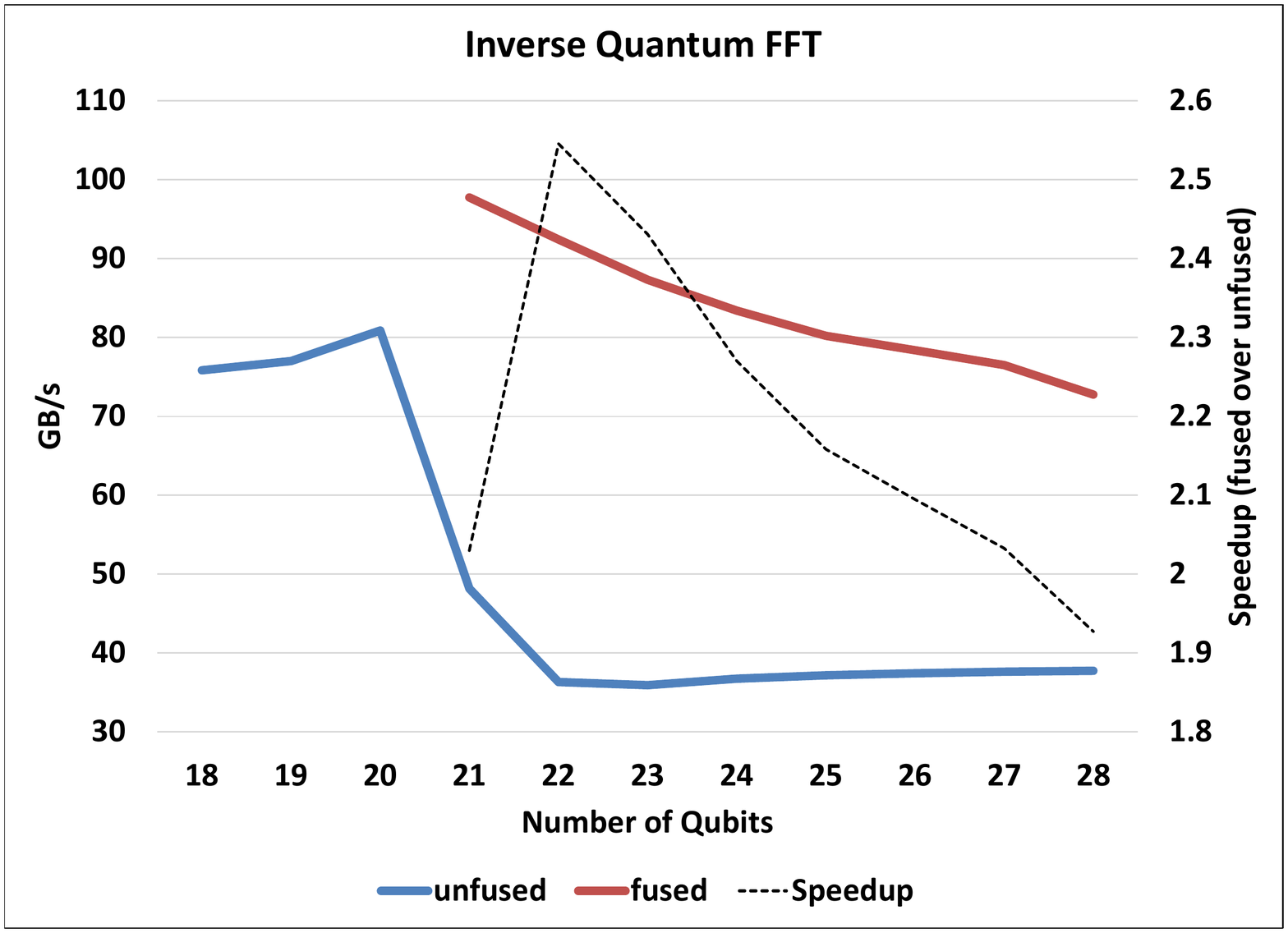}
\caption{
Performance of gate fusion optimization applied to Inverse Quantum Fourier Transform (IQFT).
}
\label{fig:fusion}
\end{figure}

Figure~\ref{fig:fusion} shows performance of fusion optimization on a single node for the \textit{Inverse Quantum Fourier Transform} (IQFT), as the number of qubits varies between 18 and 29. IQFT on $n$-qubit quantum register, shown in Figure~\ref{fig:iqfft}, consists of $n$ stages. At stage $i$, IQFT applies Hadamard gate as well as $n-i - 1$ controlled rotation gates to qubit $i$, where control gates are controlled by qubits $i+1$, $i+1$, ..., $n$, respectively. 

IQFT naturally benefits from gate fusion optimization: all stages from $0$ to ${l_c}$ can be fused and therefore blocked in LLC. Results of applying fusion optimization to IQFT are shown in Figure~\ref{fig:fusion}, for the quantum register sizes between 18 to 29 qubits. For a given number of qubits $n$,  we report performance in terms of achieved bandwidth (GB/s), computed as an aggregate amount of memory traffic for all gates divided by the overall run-time of IQFT.  The closer the achieved bandwidth is to STREAM bandwidth of 40 GB/s, the closer IQFT performance is to memory bound. Since on our system the LLC size is 20 MB, ${l_c}=20$, and thus we can fuse all the way to the 19th stage of IQFT, independent on $n$. 

The "unfused" curve shows baseline performance without gate fusion. We see that  IQFT performance for 18, 19 and 20 qubits is $\approx 80$ GB/s per gate, which is 2x higher than the memory bandwidth of 40 GB/s. This is due to the fact that for these problem sizes, the full quantum state naturally fits into LLC and thus benefits from higher LLC bandwidth. As we increase the number of qubits, the state no longer fits into LLC, and we see a dramatic drop in performance to 40 GB/s, which is the expected STREAM bandwidth. The performance stays at 40 GB/s all the way to 29 qubits. 

The "fused" curve shows performance with gate fusion. We see that for 21 qubits we achieve 100 GB/s memory bandwidth per gate, even though the state no longer fits into LLC. Thus, we see that fusion optimization increases performance by almost $2.5\times$, compared to "nofusion." As the number of qubits increases, the performance decreases gradually, due to the fact that the number of stages which cannot be fused increases. But even for 29 qubits, we achieve a bandwidth of 70 GB/s, which is nearly $2\times$ compared to the "nonfused" version.

\begin{figure}[t!]
\centering
\includegraphics[width=1.04\columnwidth]
{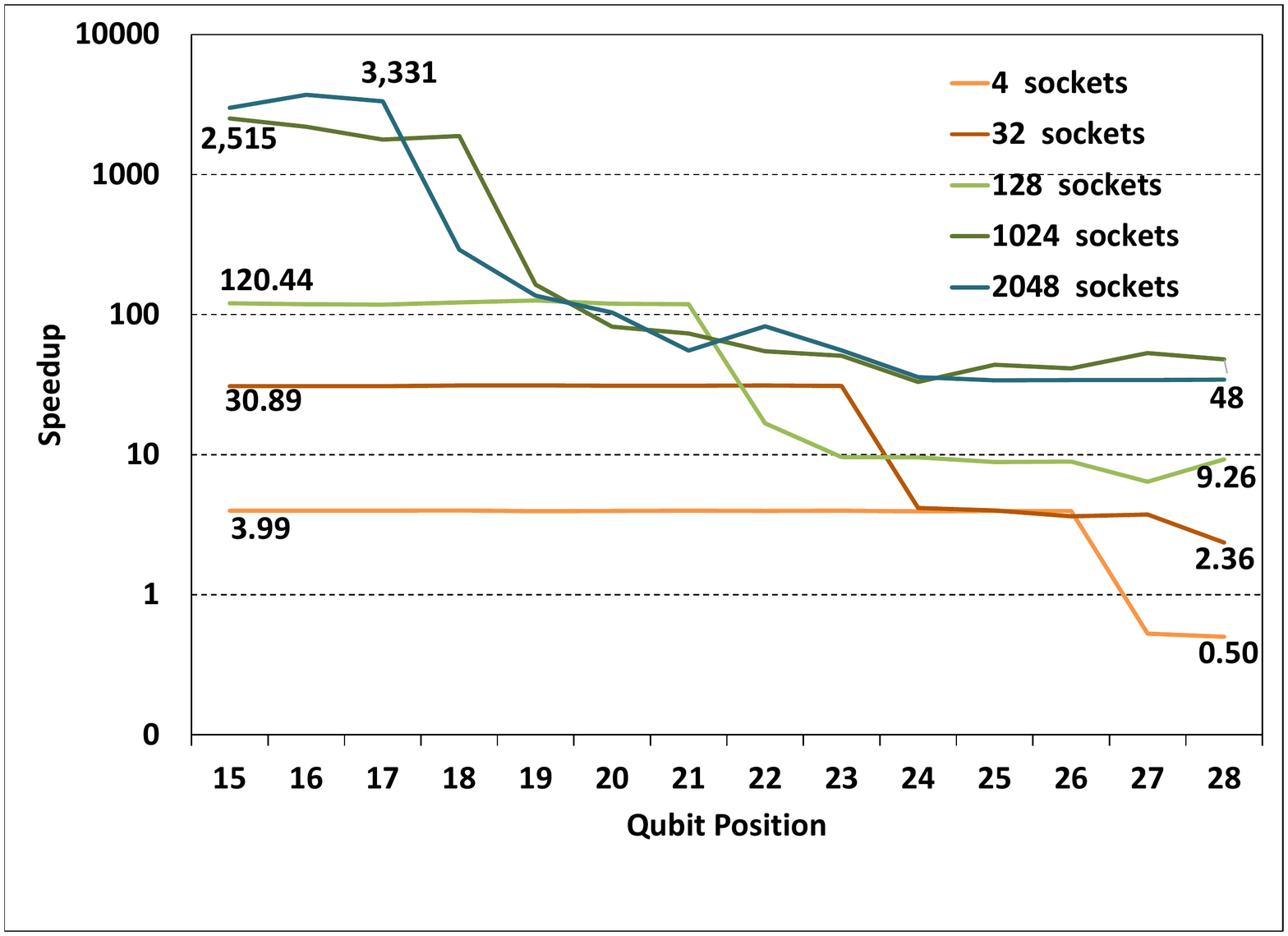}
\caption{
Strong scaling speedup over a single socket for single-qubit operation on 29 qubits. Results are shown for gate operations on qubits 15-28 for every configuration. Results for qubits 0-14 are similar to the results for qubit 15 and are omitted.
}
\label{fig:ssmultinodeperf}
\end{figure}

\subsection{Multi Node Performance}
\label{subsec:multinodeperf}

\textit{Strong Scaling:} Figure~\ref{fig:ssmultinodeperf} shows strong scaling speedup over a single socket, as we vary number of sockets from 1 to 2048. The results are shown for a single-qubit gate applied to qubits 0-28 of the 29-qubit system. The results for qubits 0-14 are similar to the results for the 15th qubit and we omit them for brevity. 

We see that on 4 sockets we achieve almost linear speedup of $4\times$ over a single socket, when the gate is applied to qubits 15 to 26, since there is no communication. Qubits 27 and 28, on the other hand, require inter-socket communication. This results in  $2\times$ slow-down, compared to a single socket, whose performance is memory bound. This is expected.  When communication is required, the performance is limited by network bandwidth, which is $\approx 7.2\times$ lower than memory bandwidth (see Section~\ref{subsec:lowerbounds}). As the result, the $7.2\times$ slowdown negates the expected $4\times$ speedup, and results in a slowdown of $0.5\times$ ($\approx 4 / 7.2$).

We observe almost linear speedup for 32 and 128 sockets when no communication is required, followed by the commensurate drop in speedup when sockets have to communicate. However, in contrast to 4-socket case, the larger number of nodes overcompensates for performance drop due to lower network bandwidth, and as the result we observe speedups of $2.4\times$ and $9.3\times$, on 32 and 128 sockets, respectively.

For 2048 sockets we observe superlinear speedup for qubits 15, 16, and 17. For example, speedup for qubit 17 is $3331\times$, an additional factor of $1.6\times$ over linear speedup of 2048. This is also expected. As we increase number of sockets, the state size per socket decreases. In particular, on 2048 sockets the state occupies only 4.2MB of memory per socket ($=2^{29+4-11}$), and thus fits into LLC. As described in Section~\ref{subsec:singlenodeperf},  LLC-bound  performance is $2\times$  higher than memory bound performance of a single node, which results in commensurate performance gains at scale.

\begin{figure}[t!]
\centering
\includegraphics[width=1.04\columnwidth]
{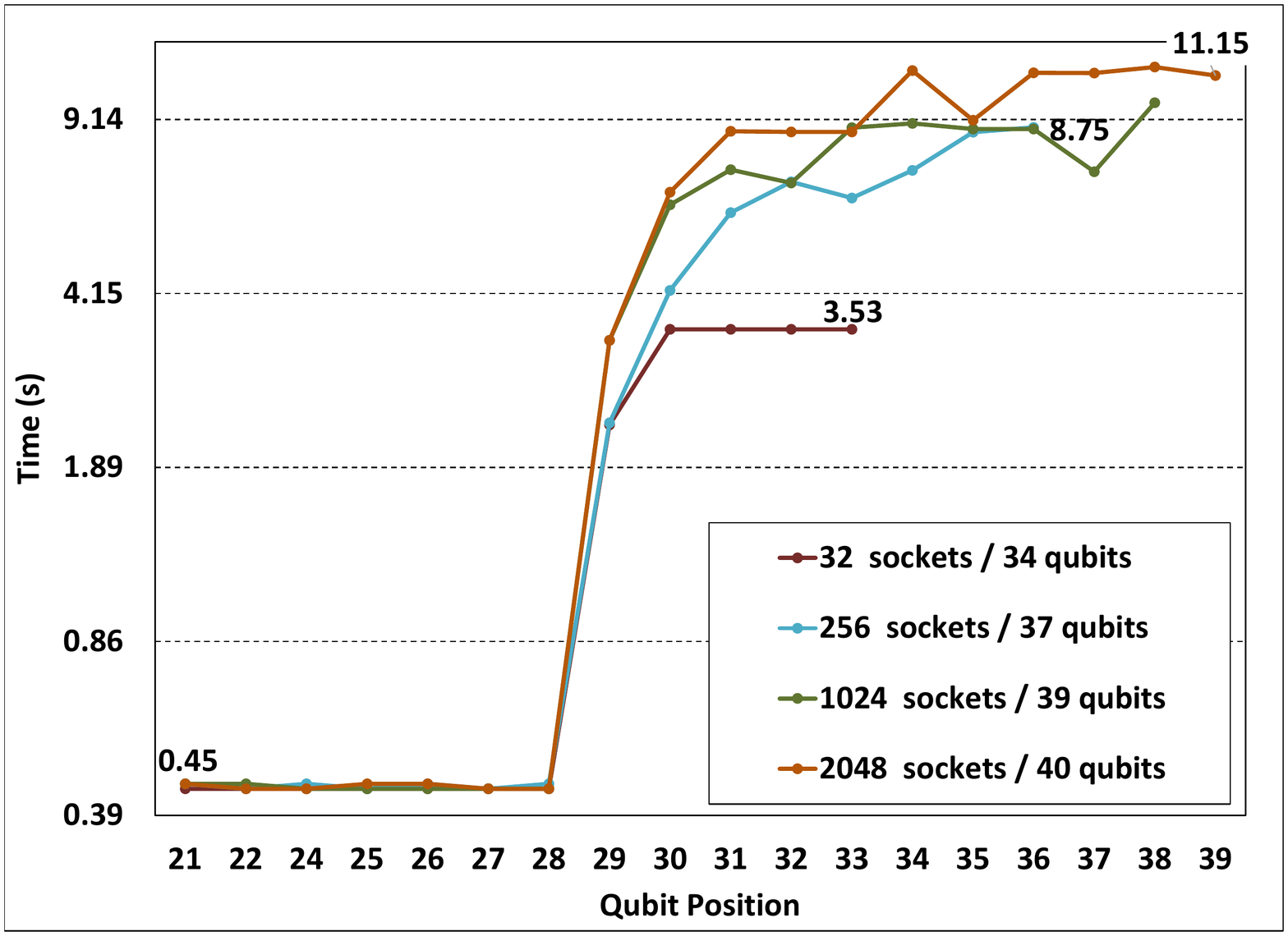}
\caption{
Multinode weak scaling of a single-qubit operation. Results are shown for gate operation on qubits $21-39$. Results for qubits $0-20$ are similar to the results for qubit $21$ and are thus omitted. }
\label{fig:wsmultinodeperf}
\end{figure}

\textit{Weak Scaling:} Figure~\ref{fig:wsmultinodeperf} shows a single-qubit gate operation on multiple nodes. We report time per gate for 32, 256, 1K, and 2K sockets, which enable simulating quantum systems with 32, 37, 39, and 40 qubits, respectively. Note this is a weak scaling experiment. Specifically, we fix the local state vector to use maximum amount of memory available on a socket. As we increase number of qubits, we also use more sockets, and as the result the size of local state vector on a socket remains the same. 

Gate operations applied to qubits $0-29$ require no communication for all four quantum systems, and achieve the performance of $\approx 0.44s$ per gate, which is very close to the memory bound of 0.43 seconds (Table~\ref{table:perfexpectations}, Case 1).  Gates applied to higher qubits require communication. For the 32-node configuration we consistently see $\approx3.53$s per gate which is within $88\%$ from network bound of 3.12 seconds (Table~\ref{table:perfexpectations}, Case 2). As the number of nodes increases, time per gate also increases, compared to the network bound. For example, gate operations applied to qubit 35 on 256-nodes configuration takes 8.7 second, which is $2.5\times$ increase, compared to network bound.  There are two reasons for such steep increase in time per gate. First, this is due to network contention.  The Stampede system uses a two-level Clos fat-tree topology~\cite{Clos} with 20 nodes (40 sockets) to the first-level switches. As a result, sockets which are more than 40 sockets apart will communicate via second-level switches. The second level switches have lower bandwidth than first level switches, thus resulting in contention and increase in run-time. Simulations using  256, 1K and 2K sockets  span multiple first-level switches. This requires communicating via second-level switched, and thus results in increased contention among communicating sockets\footnote{With topology-aware job placement, traffic on second-level switches can be prevented for gate operations on lower-order qubits that require communication. However, on a busy HPC cluster, job scheduler typically tries to maximize overall throughput, rather than individual job latency. As a result, it assigns jobs to the available CPUs, giving only secondary consideration to the job placement that minimizes the contention.}.  In addition, there is interference with other jobs running on the system at the same time, as observed by some time variability between runs in our experiments.  In the worst case, for 2K nodes, time per gate goes up to 11.15 seconds, which corresponds to $3.5\times$ ($11.15 / 3.12$) increase in time per gate, compared to network bound.

\begin{figure}
\centering
\includegraphics[width=1.0\columnwidth]
{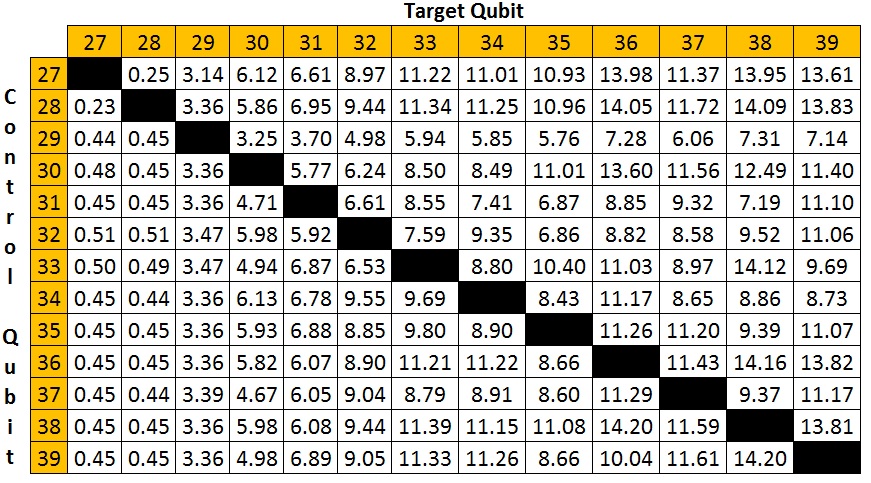}
    \vspace{0.1mm}
\caption{
Time per controlled qubit gate using 40 qubits on 1,024 compute nodes.
}
\label{fig:multinodeCG}
\end{figure}

Finally, Figure~\ref{fig:multinodeCG} shows time per controlled gate operation for a 40 qubit circuit, simulated on 2K sockets. We only present the results for values of control and target qubits in the range of $27-39$ (see Figure~\ref{fig:cgheatmap} for other combinations of $c$ and $t$). The results follow general trends similar to a single-qubit operation, discussed above. For example, Case 4 in Table~\ref{table:perfexpectations} shows the memory bound of 0.43 seconds when $t < n, c < n$ . This matches the second  columns of the table ($t=28$ and $c$ in the range $29-39$), as expected. Similar to single-qubit operation, when communication is required, the time per controlled gate can go as high as 14 seconds, which is a $4.5\times$ increase compared to network bound (Table~\ref{table:perfexpectations}).

\subsection{Performance of QFT}
\label{subsec:quantumfft}

\begin{table}[]
\centering

\begin{tabular}{|c|c|c|c|}
\hline
  &   & \textbf{Total } & \textbf{Time per} \\
\textbf{nqubits} & \textbf{ngates} & \textbf{Time (s)} & \textbf{Gate (s)} \\ \hline
29               & 435                  & 116.6               & 0.27                 \\ \hline
30               & 465                  & 141.6               & 0.30                 \\ \hline
31               & 496                  & 167.9               & 0.34                 \\ \hline
32               & 528                  & 200.9               & 0.38                 \\ \hline
33               & 561                  & 245.0               & 0.44                 \\ \hline
34               & 595                  & 297.4               & 0.50                 \\ \hline
35               & 630                  & 339.7               & 0.54                 \\ \hline
36               & 666                  & 445.4               & 0.67                 \\ \hline
37               & 703                  & 540.2               & 0.77                 \\ \hline
38               & 741                  & 643.8               & 0.87                 \\ \hline
39               & 780                  & 766.1               & 0.98                 \\ \hline
40               & 820                  & 997.2               & 1.22                 \\ \hline
\end{tabular}
\vspace{2mm}
\caption{Performance of Quantum Fourier Transform (QFT) for 29-40 qubits. The second column shows total number of gates in a single QFT call. The third column shows total time per single QFT call. Column four shows average time per gate.}
\label{table:qfft}
\end{table}

\begin{sloppypar}
Finally,  we report the performance of \textit{Quantum Fourier Transform} (QFT). QFT is fundamental kernel of many quantum algorithms, such as Shor's algorithm for factoring~\cite{Shor:1997:PAP:264393.264406}, the quantum phase estimation algorithm for estimating the eigenvalues of a unitary operator~\cite{DBLP:journals/eccc/ECCC-TR96-003}, and algorithms for the hidden subgroup problem~\cite{2004quant.ph.11037L}. The QFT circuit is similar to the one for Inverse QFT (see Figure~\ref{fig:iqfft}), except the order of qubits is reversed.  
\end{sloppypar}

Table~\ref{table:qfft} shows the performance of QFT as the number of qubits varies from 29 to 40. Gate fusion was not used in these experiments. Note this is also a weak scaling experiment, as the size of the local state vector per node is fixed to be $2^{29}$ complex amplitudes. We see that total QFT time varies from 116 seconds for 29 qubits up to 997 seconds for 40 qubits. This $9\times$ increase in run time is due to the fact that some of the gate operations become network bound, as was explained in previous section. On average, for 40 qubits, each QFT gate operations takes $\approx1.22$ seconds, as shown in the last column of Table~\ref{table:qfft}. 

Understanding  run-time requirements of a quantum algorithm is important, as it allows one to gauge the circuit complexity that can be simulated on an HPC system. For example, on Stampede cluster, a single user application is limited to a maximum run-time of 24 hours. For a 40-qubit system, this would allow $\approx86$ ($24\times3600 / 997$) calls to QFT for the total of $\approx70,000$ quantum gates.

\section{Future directions}
\label{sec:futurework}

The practical limit on the size of the quantum system that can be simulated in the next six years is 49 qubits, and can not be overcome for general-purpose circuit simulations, due to exponentially large memory requirements for storing  the entire state vector.

\begin{sloppypar}
The performance of a quantum simulator is limited by memory and especially network bandwidth. Recently introduced high bandwidth memory (HBM) delivers up to an order of magnitude more bandwidth than more traditional DRAM\footnote{HBM capacity is generally $4-8$ times lower than traditional DRAM}~\cite{HBM} and is gaining wide traction in commercial computer systems. Network bandwidth has also been improving, but at a more modest rate of $26\%$ per year. More concretely, an upcoming NERSC Cori Phase-2 system, a Cray system based on the second generation of Intel\textregistered  Xeon Phi\texttrademark  Product Family, will have an on-package, high-bandwidth memory, up to 16GB capacity and  $>400$ GB/s bandwidth ~\cite{Cori2}. The system will also employ the Cray Aries high speed dragonfly topology interconnect. These technological trends will result in performance improvement of quantum simulators.
\end{sloppypar}

However, the network contention challenge observed in the Stampede system will be inherent to low-diameter multi-level  networks, such as dragonfly~\cite{Kim:2008:THD:1394608.1382129}, widely believed to be scalable topology for exascale systems. On such networks, when gate operations affect high-order qubits, communicating processors are separated by a longer stride, which will cause congestion on the global links. This limits the overall performance of the simulator. 

Major opportunities for further accelerating \textit{q}H\textit{i}PSTER comes from communication-avoiding approaches. These approaches combine ideas of cache blocking, described in Section~\ref{subsec:fusion}, and state reordering. Cache blocking using gate fusion reduces the amount of memory traffic, but the fusion opportunities are circuit-specific and limited in scope. State reordering, on the other hand, exposes additional fusion opportunities, as described in~\cite{LIQUi}. Specifically, qubit reordering permutes the state in such a way that high-order qubits, which require inter-node communication,  become low-order qubits, thus avoiding communication for quantum gates that operate on these qubits. The main challenge is to maximize reordering opportunities for a given quantum circuit, while minimizing computational overhead of reordering. Lastly, we note that reordering optimization may allow for storing the state vector in a higher capacity secondary device, such as disk, while reducing the cost of data transfers to and from main memory. In principle, this could enable quantum simulations with more than 49 qubits.

\section{Acknowledgments}
\begin{sloppypar}
The authors also acknowledge the Texas Advanced Computing Center (TACC) at The University of Texas at Austin for providing HPC resources that have contributed to the research results reported within this paper. URL: http://www.tacc.utexas.edu.
\end{sloppypar}

\bibliographystyle{abbrv}  
\bibliography{references}

\end{document}